# Secure and Energy Efficient Remote Monitoring Technique (SERMT) for Smart Grid

Sohini Roy, Arunabha Sen

Arizona State University, Tempe AZ 85281, USA
sroy39@asu.edu, asen@asu.edu

**Abstract.** Monitoring and automation of the critical infrastructures like the power grid is improvised by the support of an efficient and secure communication network. Due to the low cost, low power profile, dynamic nature, improved accuracy and scalability, wireless sensor networks (WSN) became an attractive choice for the Information and Communication Technology (ICT) system of the smart grid. However, the energy efficiency and security of WSN depends highly on the network design and routing scheme. In this paper, a WSN based Secure and Energy Efficient Remote Monitoring Technique (SERMT) is proposed by demonstrating a WSN based ICT network model for pervasive monitoring of the generation and transmission part of the power network in a smart grid system. The performance of the proposed network model designed for a smart grid of IEEE 118-Bus system coupled with the secure routing technique is tested during cyber-attacks by means of NS2 and the simulation results indicate that it performs better than existing smart grid monitoring methods like Lo-ADI [1] with respect to packet drop count and throughput.

## 1 Introduction

A sustainable lifestyle for human race is reliant upon critical infrastructures like power network, communication network, water supply system, transportation system etc. Therefore, pervasive monitoring [2] and control over such critical systems is very important. It is to be noted that only an improved Information and Communication Technology (ICT) can transform every such system into a smart system and thus can form a smart city. Smart grids are an inherent part of a smart city and it is obtained by incorporating features like full-duplex communication between the ICT entities, automated metering in the smart homes, power distribution automation, and above all intelligent decision making by means of pervasive monitoring of the power system and securing its stability. Therefore, it is beyond any question that the ICT network of a smart grid must be accurate, scalable, and secure enough to instantly identify any kind of abnormal behavior in the entities of the power network, securely communicate that information to the control center (CC) and thereby help in taking necessary and timely action to ensure uninterrupted power supply.

Although, designing of a robust ICT network for smart grid has become a boiling topic of research, the best suited ICT design for a smart grid is still not very clear. In [3] only a crude idea of the design of a joint power-communication network is given using a test system consisting of 14 buses. While, the authors of [4], have come up with a realistic design of the ICT for smart grid by taking help from a power utility in the U.S. Southwest; their ICT system relies completely on wired channels that either use SONET-over-Ethernet or Ethernet-over-Dense Wavelength Division Multiplexing. A completely wired ICT system is neither cost effective nor energy saving. Every communication entity in [4] draws power and thus a huge amount of power is devoted for monitoring the power network itself. Moreover, addition/isolation of ICT entities for hardening purpose or fault tolerance/during a failure or a security threat is extremely difficult and costly in a wired system.

Smart sensor nodes like Phasor Measurement Units (PMUs) are popular in remote monitoring system for smart grid. Power generation and transmission, power quality, equipment fitness, load capacity of equipment and load balancing in the grid can also be monitored by data sensing techniques. As a result, wireless sensor networks (WSNs) are currently acquiring attention of the researchers for the purpose of designing the ICT network for a smart grid system. WSNs are comprised of low powered sensor nodes with easy installation process, lesser maintenance requirement, low installation cost, low power profile, high accuracy and scalability. All these have convinced the researchers that WSNs are a very good choice for the designing of the ICT of a smart grid. Therefore, in this paper, a new design of the ICT network for a smart grid is proposed using WSN.

Despite all these advantages, the most common drawback of a sensor node is that it is battery powered and battery replacement is difficult. As a result, energy conservation becomes important. In Secure and Energy Efficient Remote Monitoring Technique (SERMT) energy efficiency is obtained by energy aware routing and use of more expensive rechargeable Energy Harvesting Sensor Nodes (EHSNs) for MU and PMU data transfer respectively. Also, the sensor nodes as well as the wireless channels are vulnerable to common cyber-attacks [5] like sinkhole, wormhole, Sybil attack, packet eavesdropping, flooding attack and most importantly false data injection attack by means of node compromise. SERMT aims at securing the data transfer together with energy conservation by means of light weight security protocols used in [6] like Elliptic-Curve-Public-Key Cryptography (ECC), Elliptic-Curve-Diffie-Helman Key exchange scheme (ECDH), Nested Hashed Message Authentication Codes (NH-MAC) and RC5 symmetric cypher.

The rest of the paper is structured as follows. Section 2 gives an overview of the ICT network setup phase for SERMT. A realistic risk model for the SERMT network design is given in section 3 of this paper. Mitigation of the threats discussed in section 3 by adopting a secure and energy efficient routing technique is explained in section 4. Section 5 does performance analysis of the proposed scheme SERMT by means of comparing the simulation results with an existing secure remote monitoring technique for power grid named as Lo-ADI [1]. Sections 6 concludes the paper and discusses the scope for future works.

## 2     Overview of the ICT network setup phase for SERMT

In order to provide a reliable remote monitoring for the smart grid, a generic ICT network design is proposed by SERMT that can be applied on any given power network. In this section, the network setup phase for SERMT is divided into four steps. Each step is illustrated with the help of IEEE 14-Bus system as the power network.

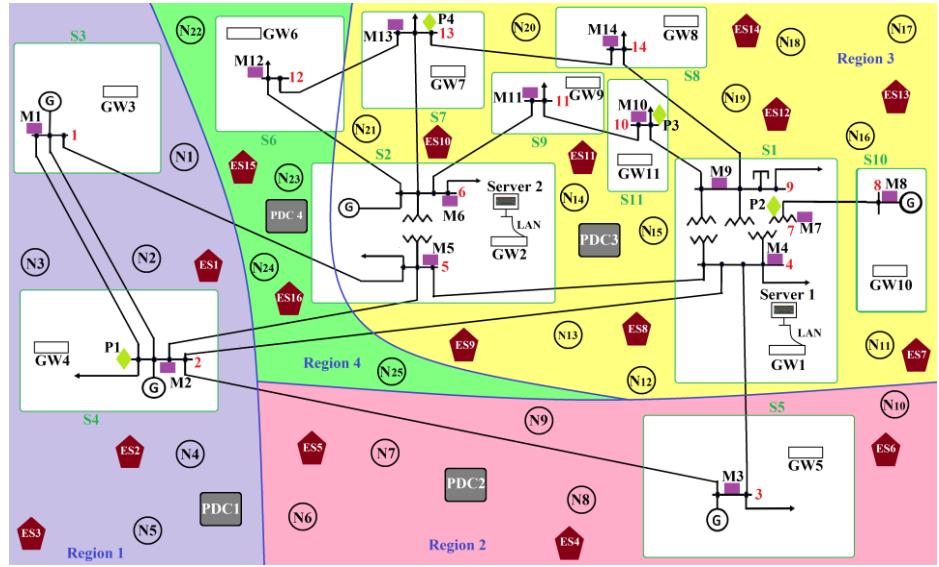

**Fig. 1.** Region division and sensor node placement in a smart grid of IEEE 14-Bus system

In the first step, a given power grid is divided into several substations. The buses connected by transformers are placed in one substation with the assumption that transformers remain within a substation only [7]. This process is repeated until no transformers are left. Then each bus is placed in a unique substation. In Fig.1, the IEEE 14 bus system is divided into 11 substations denoted by $S_i$ where i is the substation ID.

After the substation division, the distance between all pairs of substations ($S_i$ & $S_j$) is calculated [4]. Now, starting from a substation $S_i$ with the maximum connectivity among the substations at the network borders, all substations which are within a given distance of it, are marked as substations of a common monitoring region $R_x$. Then the next substation which is the closest to $S_i$ but beyond the given distance and which is not yet placed in a monitoring region, is selected and the same process is repeated. This process is continued till every substation is placed within a monitoring region. For example, in the IEEE 14-Bus system, substations $S_3, S_4, S_5, S_6, S_7, S_8$ and $S_{10}$ are at the borders of the smart grid area. Among them, $S_4$ has the maximum connectivity, therefore the region division starts from it. Fig 1. shows that after this process is completed, the smart grid network for IEEE 14-Bus system is divided into 4 regions.

SERMT considers two different types of smart sensors for monitoring the power network entities. The first type is the Measuring Unit (MU) based smart sensors [8].

The MU-based smart sensors have a sensing unit consisting of sensors like Current Transformer (CT) and Voltage Transformer (VT) and a merging unit for performing signal amplification, conditioning and analog-digital conversion. There is an internal clock in the merging unit that can timestamp every sample and thereby time synchronize them with an external UTC source. The MU-based sensor has a network communication module which helps it to communicate with Intelligent Electronic Devices (IEDs) or applications. The second type of sensors used for SERMT is the PMU-based smart sensors. The sensing module of a PMU has sensors like voltage and current sensor which provide analog inputs for the PMU. Current and voltage phases are sampled at a rate of 48 samples per cycle in a PMU using a common time source for synchronization. It also has an internal clock and a processing module responsible for signal conditioning, timestamped analog-digital signal conversion and computing voltage and current phasors to produce a synchronized phasor and frequency.

In this step, MU-based sensors are placed at every bus in the power network. PMU-based sensors are placed at some of the buses using an optimal PMU placement algorithm [7]. Low-cost, non-rechargeable battery enabled sensor nodes denoted by $N_j$ with ID $j$ are randomly dispersed across the network area. These $N_j s$ can carry the MU-based sensor data to the CCs in a power efficient manner. A phasor data concentrator (PDC), responsible for receiving accumulating PMU data from multiple PMUs, is placed at each region of the smart grid and few EHSNs denoted by $ES_j$ with $j$ as the ID are randomly deployed across the smart grid region. The idea behind the deployment of the two kinds of sensor nodes is that, the cheaper $N_j s$ will follow an event-driven hierarchical routing approach to send the MU data from the MU-based sensor nodes to the CCs; and the $ES_j s$ will always be active to accumulate synchrophasor data from the substations of each region and send the data to the local PDCs and finally to the CCs. Due to the high volume of PMU data transfer from each substation having a PMU, the sensor nodes carrying them should always be active.

In Fig.1. the MU-based sensors are denoted by $M_i$ where i is the ID of the sensor and it is same as the bus number it is placed at. The PMUs are represented by $P_i$ and they are numbered in the order they are placed. It is assumed that $N_j s$ and $ES_j s$ are deployed region wise and are numbered in the order of their deployment.

SERMT assumes that there is a main CC and a backup CC. So, even if the main CC is damaged, the backup CC can efficiently monitor the network state. The connectivity of each substation is calculated based on the number of transmission lines connecting that substation with other substations. The substation having the highest connectivity is selected as the main CC and that having the second highest connectivity is selected as the backup CC. After the CCs are selected, all substations are equipped with a router acting as a gateway for the substation servers and the servers in each CC are access points for the operator. The CC-gateways can wirelessly receive data from the sensors and PDCs and send those data to server via a wired LAN connection in control centers. In the Fig.1., $S_1$ is the main CC and $S_2$ is the backup CC.

## 3     Risk model and assumptions for the SERMT network setup

The WSN nodes are vulnerable to several attacks that are explored by researchers over the ages. In SERMT, some of the common attacks on WSN are considered. The threat model for SERMT predicts that a compromised entity of the ICT can pretend as a CC-gateway and congest the network with bogus communications to launch a flooding attack which can result in denial of service (DoS) by the ICT entities. Another common attack of the WSN is the Sybil attack [5] where a legitimate node of the network gets compromised and deceits the other nodes by providing multiple fake location information of itself. In SERMT network setup, any ICT entity except the substation entities can impose a Sybil attack and oblige other entities to forward data to an entity that does not exist in that location resulting in increased packet drop. In a node compromise attack, the attacker physically gets hold of the node and reprograms it to launch attacks like wormhole attack and packet eavesdropping attack [5]. These attacks can also be launched by an outsider's node deployed in the network. The Sinkhole attack can be launched by attackers with high quality equipment and they can inform the ICT entities about a high-quality entity in their sensing zone, so that all the other entities select it to forward the data to the CC. This node gathers all the data the CCs do not receive any information about the system state from the network.

In order to provide security to the ICT infrastructure for the smart grid, the following assumptions are made in the proposed work:

1. The substation equipment like the server and the gateway are trustworthy and it is impossible for an attacker to compromise those components. The servers authenticate themselves to the sensors or PDCs to avoid flooding attack [5].
2. Each entity of the ICT network is provided with a global key *GBK* which an attacker cannot get hold of even if the entity is compromised.
3. A unique set of elliptic curves is stored in the memory of each ICT entity for the purpose of ECC and ECDH protocols. Also, in order to achieve those mechanisms, it is assumed that any pair of entities in the network agrees upon all the domain parameters of the elliptic curves stored in them.
4. The two CC-servers generate two separate one-way hash chains using SHA-1 hash function [6]. The starting key for each chain is randomly selected by the each of the servers. All the ICT entities are provided with same hash functions and the last keys of the two chains so that they can authenticate the control messages from the control center servers and avoid a Sinkhole attack.
5. Each ICT entity in the network is aware of its own ID, location information, region ID and ID of the servers.
6. All the PDCs can communicate with other PDCs in the neighboring regions.

## 4     Secure and Energy Efficient Routing Scheme of SERMT

The goal of the ICT network for a smart grid is to securely transmit the sensed data from the sensors to the control centers and help in remote monitoring of the power grid. In order to achieve this with the help of a WSN based ICT network, a secure and energy efficient routing technique should be followed. In this section, the secure routing scheme of SERMT is divided into 4 modules.

### 4.1 Module 1: Initial trust value estimation of the ICT entities

In the first module of SERMT, the main control center server ($Server_1$) sends a number of test messages to the sensor nodes and PDC in its own region to calculate their Trust value ($TV_i$) . The message format is: $Server_1 \rightarrow *$ : $MSG||K_x||HMAC(GBK;MSG||Server_1ID||K_x)$ where * denotes a set of ICT entities, MSG is the test message, $K_x$ is a random key from the sequence of keys generated by the hash function and this key is never used again. A Hash based Message Authentication Code (HMAC) is generated over the MSG, server ID and the key using the shared global key GBK and appended with the message, so that any non-legitimate node which do not have the GBK and the server ID cannot separate the HMAC from the original message and therefore cannot overhear the message from the server to the other entities. The $TV_i$ is calculated by the server for each entity i using eq.1.

$$TV_i = \frac{MSG_{delivered}}{MSG_{sent}} * 100 \qquad (1)$$

Entities having a $TV_i$ of more than 40% are selected as trusted nodes and any legitimate node having a lower $TV_i$ are added to the threat list of the server. Now, the node $X_i$ with the highest $TV_i$ in that region is selected by the CC-server to forward a request message to the nodes closest to it in the adjacent regions. The format of the request message is: $Server_1 \rightarrow X_i: RQM||||K_x||HMAC(GBK;MSG||Server_1ID||K_x)$ where RQM is the request to calculate $TV_i$, $K_x$ is a new random key from the key chain and HMAC is generated and appended with the message. First, the $TV_i$ of the nodes of adjacent regions closest to $X_i$ are calculated by $X_i$ in the same manner. If those nodes have a $TV_i$ greater than 40% then the RQM is forwarded to them. These nodes now calculate the $TV_i$ of the nodes in their respective regions and send a list of blocked node IDs having low $TV_i$ to the CC-server via $X_i$. The message format is: $e_i \rightarrow Server_1: Blocked\_List||e_iID||HMAC(Blocked\_List||e_iID)$ . The $Server_1$ updates the threat list using this data. This process is repeated till the threat nodes of each region are identified by $Server_1$. $Server_1$ shares the threat list with $Server_2$ via a shortest path consisting of nodes in the trusted list. After both the servers get the list of trusted nodes in the network, the trust value of each node is shared with the substation gateways by $Server_1$, via a shortest path to the substation consisting of the most trusted nodes. This whole process is repeated at a regular time interval by alternate CC-servers. Each time the new trust values are assigned to the ICT entities, both the servers compute a public-private key pair $(U_{Server_i}, v_{Server_i})$ and share their public key with all the entities in the network by broadcasting.

## 4.2 Module 2: Data sensing by the smart sensor nodes and data forwarding by substation gateways

The next phase of SERMT is the data sensing phase. In this phase, the smart sensors collect data from the power entities and forward that to the substation gateways. The substation gateway is assumed to be trusted. The CC-gateways forward the data directly to the connected servers, but they still need to send data to the other CC. The gateways work differently for sending MU-based and PMU-based sensor data to the CCs. In order to send the MU-based sensor data, the gateways send a forward request message to each $j^{th}$ node ($N_j$) of that region at one-hop distance from it. The format of the request is: $GW_i \rightarrow N_j: forw\_RQM||GW_iID||HMAC(GBK; forw\_RQM||GW_iID)$. Each $N_j$ receiving the forw_RQM from the gateway sends back an acknowledgement (ACK) appended with information like their remaining battery power ($BP_j$) and its connectivity with other nodes in the same region ($C_j$) to the sender in the format: $N_j \rightarrow G_i: ACK ||BP_j||C_j||HMAC(GBK; ACK||BP_j||C_j)$. In case a node receives forw_RQM from multiple gateways in that region, it selects the one closest to it. The substation gateways calculate a data forwarding score $DF_j$ for all the nodes sending back ACK to it using eq. 2.

$$DF_j = BP_j * TV_j * C_j \qquad (2)$$

Now, each gateway $GW_i$ generates a key pair having a public and a private key $(U_{GW_i}, v_{GW_i})$ using ECC and forward the public key $U_{GW_i}$ to the $j^{th}$ sensor node having the highest $DF_j$ calculated by itself using the message format: $GW_i \rightarrow N_j: U_{GW_i}||GW_iID||HMAC(GBK; U_{GW_i}||GW_iID)$. The $N_j$ receiving the public key from the $GW_i$ also generates a private-public key pair $(v_{N_j}, U_{N_j})$ using ECC and send back the public key and its ID in the similar format to $GW_i$. $GW_i$ now computes a point in the elliptic curve $(x_k, y_k) = v_{GW_i}.U_{N_j}$ and $N_j$ computes a point $(x_k, y_k) = v_{N_j}.U_{GW_i}$ where $x_k$, the x coordinate of the computed point becomes their shared secret. This is ECDH key exchange scheme where the shared secret calculated by both parties is same– $v_{GW_i}.U_{N_j} = v_{GW_i}.v_{N_j}.G = v_{N_j}.v_{GW_i}.G = v_{N_j}.U_{GW_i}$; where G is the generator of the elliptic curve. This shared secret $x_k$ is the temporary key for exchanging encrypted data between the $GW_i$ and $N_j$. $GW_i$ now encrypts the aggregated MU data from the buses of that substation using RC5 symmetric cypher with $x_k$ and forwards that to $N_j$ by using NH-MAC in the format: $GW_i \rightarrow N_j: EMD||GW_iID||HMAC(GBK; HMAC(x_k; EMD||GW_iID))$. The gateways keep on forwarding the encrypted MU data to the same node until the servers initiate a new round of module 1.

At the same time, the PMUs also sense data from the power entities and send that to the corresponding substation gateway. Each PMU containing substation gateway now forwards forw_RQM to the EHSNs of that region within its range. In the similar way, the $ES_j$s receiving the forw_RQM sends back ACK to the gateway with only the connectivity information. Now, the gateway recalculates the $TV_j$ for each $ES_j$ like module 1 and selects the node with the highest current $TV_j$. The same key exchange scheme is established with the most trusted $ES_j$ as it was done with the $N_j$s and the RC5 encrypted PMU data is forwarded to $ES_j$ in the similar format with NH-MAC. $GW_i$s recalculate the $TV_j$ of the $ES_j s$ at regular intervals to repeat this process of key and data exchange and do not wait for the servers to estimate $TV_j$ of the $ES_j s$. However, if a server reinitiates module 1, all the substation gateways wait till the process is complete and they are updated with the new trust values of all the nodes in their region. The CC-gateways also use the same process to send data to the other CC.

### 4.3   Module 3: Data forwarding to the control centers

In this module, two separate routing techniques are followed by the $N_j s$ and $ES_j s$ to forward data to the CCs. $N_j s$ receiving data from the $GW_i s$ first decrypt them using the same temporary key $x_k$. Then, a join request (join_RQM) is broadcasted in the format: $N_j \rightarrow *: join\_RQM || N_j ID || HMAC(GBK; join\_RQM || N_j ID)$ to all other $Ns$ in that region by that $N_j$. Malicious nodes present in the network receiving the join_RQM cannot decode the HMAC as they do not have the GBK and thereby they cannot join any legitimate data-carrying $N_j$. Each $N_j$ receiving ACK from other $Ns$ provide a cluster ID to each of the $Ns$ in the format RegionID||Timestamp_for_current_$TV_i$ ||List_of_SubstationIDs. The server assigned timestamp value for current $TV_i$ is same for all the nodes in a common region and the list of substation IDs denote the data carried from which of the substations in that cluster. Therefore, all the nodes getting the common cluster ID are considered as nodes of a cluster. Multiple clusters can be formed in each region of the smart grid.

After the cluster formation in every region, a cluster head (CH) is selected for each cluster using a candidate value ($CV_j$) calculated using eq.3. by each node j in the cluster and shared with all other nodes in the cluster.

$$CV_j = BP_j * TV_j * Cn_j \qquad (3)$$

In eq.3. $BP_j$ is the remaining battery power of the node, $TV_j$ is the current trust value of the node and $Cn_j$ is the connectivity of that node with the nodes of other adjacent regions. The $Cn_j$ of a node is determined based on the number of nodes in other regions that are within the sensing radius of the node j. Now, the CH generates a public-private key pair like $GW_i$ and share the public key with each of the nodes in the cluster. The cluster member nodes (CMNs) also generate the key pair, send their public key to the CH and compute their own $x_k$ using ECDH. Each CMN now forwards its data to the CH by encrypting the data with the $x_k$ using RC5 and generating NH-MAC just like each $GW_i$ did. The CHs after receiving the data from the CMNs, decrypt them with the $x_k$ for that CMN. Now each of the CHs aggregate the data obtained from the member nodes and encrypt that with the public key $(U_{Server_1})$ of the main CC-server using ECC based public key cryptography. In case, the main CC fails, the network is informed about that and all data are forwarded to the backup CC.

Now, a CH of the main CC region, directly forwards the encrypted-aggregated data to the main CC-gateway. Other CHs use a modified version of Dijkstra's shortest path algorithm to send the aggregated data to the main CC-gateway. In this algorithm, the weight values $(W_{ij})$ of the link between two nodes i and j are determined by eq.4.

$$W_{ij} = {D_{ij}}/{(BP_j * TV_j)} \qquad (4)$$

In eq.4. $D_{ij}$ stands for the distance between the sender i and the receiver j, $BP_j$ is the remaining battery and $TV_j$ is the trust value of the node j. The same route is followed by the nodes for sending data to the main CC until new nodes are selected by $GW_i s$ for data forwarding. The main CC-gateway generates the HMAC over the received message to check its authenticity and if a match is found, the data is forwarded to $Server_1$. $Server_1$ periodically sends backup to the $Server_2$ via the shortest path from $Server_1$ to $Server_2$, determined using the same modified Dijkstra's algorithm.

In order to forward the PMU-based sensor data to the CCs, each data-carrying $ES_j$ node checks if the PDC of that region is adjacent to it. Each $ES_j$ is updated with the location information and the current $TV_j$ of other $ES_j s$ and the PDC in that region by the $GW_i s$ of that region. If a PDC is adjacent to an $ES_j$, then the $ES_j$ first decrypts the data and generate a public-private key pair to share the public key with the PDC. The PDC also does the same and share back its public key with the $ES_j$ to complete the ECDH key exchange. The node $ES_j$ now forward the encrypted NH-MAC-ed data to it. $ES_j s$ which are not near a PDC, select the shortest path consisting of the most trusted $ESs$ from itself to the PDC in that region to send their public key and the same route is followed by the PDC to send back its own public key to the $ES_j$. Now, the encrypted data is forwarded through the same path to the PDC. The PDC of each region acts in the similar way as a CH. It decrypts the data from each $ES_j$ using the unique $x_k$ for that node and aggregates the whole data. A shortest path consisting of PDCs only, from each PDC of the network to both the control center gateways are calculated only once and is never repeated until a PDC of a region is replaced or a PDC fails. In case, the $TV_i$ of a PDC drops below the threshold, the $Server_1$ adds it to the blocked list and selects the most trusted $ES_j$ node to perform the function of a PDC, until the actual PDC is replaced by a new one or its fault is diagnosed and recovered by the operator. Now, the aggregated data is encrypted using the public key of the $Server_1$ by each PDC and forwarded to the main CC-gateway. The same process is repeated for $Server_2$. Both the CC-gateways authenticate the received data and forward it to their connected servers for evaluation.

The energy consumed by both the N nodes and ES nodes in transmitting a message $(E_T)$ is the summation of energy consumed by their power amplifier $(E_A)$ and it is a function of distance from which the message is sent, baseband DSP circuit $(E_B)$ and front-end circuit $(E_F)$. The energy consumed in receiving a message $(E_R)$ is the summation of $E_B$, $E_F$ and the energy consumed by Low-Noise Amplifier $E_L$.

## 5 Simulation Results

In this section, a smart grid network of IEEE 118-Bus system is considered. The total network region is divided into 8 regions and the power grid is divided into 107 substations. Substation 61 is selected as the main CC and it consists of 3 buses–68,69 and 116. Substation 16, consisting of buses–17 and 30, is selected as the backup CC. The main CC is in region 1 and the backup CC is in region 2 of the smart grid. In order to analyze the performance of SERMT in this network setup, a total of 500 N type nodes, 300 ES type nodes and 8 PDCs (high quality EHSNs with longer communication range) are deployed in the network area and simulation is performed using NS2.29 network simulator. The simulation results are compared with the Lo-ADI [1] technique that incorporates a secure token exchange mechanism among the WSN nodes. The simulation parameters are given in table.1.

**Table 1.** Parameter list for simulation

| Parameter | Description |
| --- | --- |
| Operating system | Fedora |
| Simulator | NS2.29 |
| Compiler | TCL |
| Rechargeable battery for EHSNs | NiMH |
| Battery capacity of EHSNs | 2000(mAh) |
| Initial battery power for all nodes | 150 (mAh) |

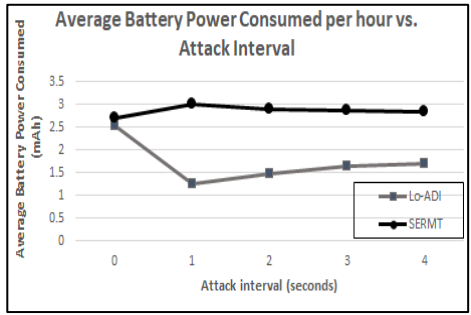 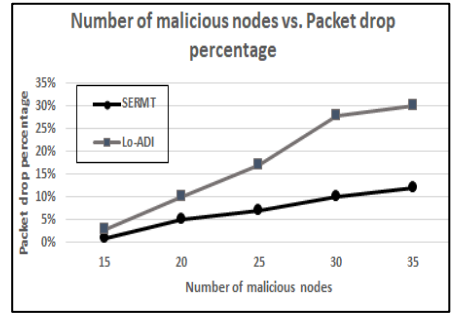

**Fig. 2.** Avg. BP consumed vs. Attack interval  **Fig. 3.** Number of malicious nodes vs. packet drop percentage

In Fig.2, the average battery power consumed per hour by all types of sensor nodes is shown for different attack intervals in both Lo-ADI and SERMT. It is observed that the average battery power consumption of SERMT is a little higher than that of Lo-ADI and this battery power consumption does not decrease if the attack interval is 1 second as in Lo-ADI due to connection revoke and packet drops, rather its battery power consumption increase if the attack interval is less as SERMT immediately adds the malicious node into its threat list and resends the packet to some other trusted node. Moreover, due to the presence of ESs, the average residual energy in the network will always be more than that of Lo-ADI. In Fig.3, it is proved that even though

SERMT consumes more power than that of Lo-ADI because of using several power consuming security mechanism computations and key exchange protocols, it wins in terms of offering security to the smart grid. Fig.3. shows, percentage of packets dropped in presence of different number of malicious nodes in the network and it is observed that in SERMT, the packet drop percentage is much lesser than that of Lo-ADI even when there are 35 malicious nodes in the network.

## 6      Conclusion and Future Works

The region based remote monitoring adopted by SERMT helps in easy identification of a failure in the power grid or an attack in the communication network of the smart grid. Security of sensor data is of utmost importance in the smart grid as any alteration of the data can lead to wrong decision by the operator and this might badly harm the smart grid. Data privacy is obtained in SERMT by the encryption/decryption mechanisms, data integrity is verified by the hashed message authentication and authentication of data is done by the one-way hash chain, therefore all the data security measures are addressed by SERMT. One drawback of SERMT is that it cannot deliver the sensor data from the sensors to the control centers in a very short time due to the security measures it follows and this can be explored in future research how the time delay be reduced in a secure communication technique for smart grid. Designing a threat model with attacks on MUs, PMUs, gateways or servers and analyzing the effect of cyber-attacks on power grid can be another direction of future work.